\documentclass[aps,prb,twocolumn,floatfix,superscriptaddress]{revtex4}
\usepackage{graphicx}
\usepackage[latin1]{inputenc}
\usepackage{dcolumn}
\usepackage{bm}

\begin{document}

\title{A diffraction-compensating 0--25 ns free space terahertz delay line for coherent quantum control}

\author{D. G. Allen} 
\affiliation{Department of Physics, University of California,
Santa Barbara, California 93106}
\affiliation{Institute for
Quantum and Complex Dynamics, University of California, Santa
Barbara, California 93106}
\author{S. Takahashi} 
\affiliation{Institute for Quantum and Complex Dynamics,
University of California, Santa Barbara, California 93106}
\author{G. Ramian} 
\affiliation{Institute for Quantum and Complex Dynamics,
University of California, Santa Barbara, California 93106}
\author{M. S. Sherwin}\email{sherwin@physics.ucsb.edu}
\affiliation{Department of Physics, University of California,
Santa Barbara, California 93106} \affiliation{Institute for
Quantum and Complex Dynamics, University of California, Santa
Barbara, California 93106}
\author{L. Persechini} 
\affiliation{School of Physics, Trinity College Dublin, College
Green, Dublin 2, Ireland}
\date{\today}

\begin{abstract}
Free space delay lines provide pulses of variable time spacing for
optical experiments such as pump-probe spectroscopy and coherent
quantum control, including spin and photon echo techniques.
However, in the terahertz (THz) region of the spectrum, beam
divergence due to diffraction limits the useful length of
traditional free space delay lines. We present a novel
double-folded variable delay line for light in the frequency range
0.24--1.2 THz, which incorporates a symmetric arrangement of
lenses whose spacing can be adjusted to compensate for diffraction
at each delay. Scalable for use in other wavelength regimes, the
design relays an input Gaussian beam waist to the output with up
to 25 ns ($\sim 8$ m) total delay and is enclosed in a desiccated
volume of $<0.5$ m$^3$. The delay line can deliver two or three
pulses with relative amplitudes controlled via variable spacing
silicon etalon beam splitters. Beam profiles of a 0.24 THz beam
show good agreement with calculations at long delays, with
insertion loss per delay stage of $\sim3$ dB.
\end{abstract}


\maketitle The quest for higher spectral resolution, higher
sensitivity and faster time response has driven state of the art
electron paramagnetic resonance (EPR) beyond microwaves, to the
edge of the terahertz (THz) regime\cite{eprref}. At the same time,
the search for a qubit with suitably low decoherence rates and
potential scalability for quantum information processing has led
some researchers to the same technologically under-serviced
spectral range\cite{sherwinqubit,lithiumdonors}. In both fields,
pulsed quantum control techniques allow measurements of population
relaxation (via time-resolved decay of magnetization or emission),
ensemble dephasing (via decay of Rabi oscillations versus pulse
intensity or duration), and decoherence (via the decay of a spin
or photon echo). Such measurements yield information about local
environments and interactions at the nanoscale, and are being
pursued at terahertz frequencies for studies of protein conformal
dynamics using amino acid spin labels and dynamics of electrons in
quantum confined nanostructures. We note the application of these
techniques is predicated on the availability of high intensity
radiation delivered in pulses of well-defined shape, controllable
amplitude and variable time delay.

In the optical and infrared regions of the spectrum, free space
delay lines utilize a beamsplitter to direct a portion of an
optical pulse along variable path length, $d$, terminated by
retroreflecting mirror arrangement, which offsets the pulse and
returns it along a parallel trajectory. The reflected portion is
delayed relative to the incident pulse by $2d/c$, where $c$ is the
speed of light. At the other end of the spectrum, below 0.1 THz,
waveguided microwave devices are preferred over free space delays
for pulse generation.

In the intermediate range 0.1--10 THz, neither ray optics
solutions nor waveguided circuit technologies for pulse generation
are generally applicable. This is due to a conspiring
collaboration of failing material properties and the breakdown of
most of the useful approximations that make microwave and
photonics tools readily engineerable. THz sources based on
semiconductor transport including Schottky diode
multipliers\footnote{Commercially available from, e.g., Virginia
Diodes, Inc..} and quantum cascade lasers\cite{qcl} can be
modulated by microwave techniques, but do not yet provide adequate
power for most quantum control experiments. Thus, quasi-optical
sources and pulse generating techniques are required for most
nonlinear and quantum control experiments\cite{carterscience}. In
such quasioptical systems, beam diffraction and water vapor
absorption are primary concerns.

Here we describe a folded, dry atmosphere delay line which
compensates for diffraction, enabling delay paths of up to $\sim8$
m ($\sim25.6$ ns) for wavelengths as long as $1.3$ mm in compact
2x0.5 m footprint. The delay line is designed as dual use
instrument for studies of quantum dynamics of electrons confined
by impurity centers\cite{colenature}, quantum well
heterostructures and quantum dots, and as part of a high field,
240 GHz pulsed EPR (a.k.a. ESR) spectrometer under collaborative
development by the UCSB Center for Terahertz science and
Technology and the National High Magnetic Field Laboratory,
Florida\cite{injectionlocking}. The delay line delivers two
controllable amplitude pulses of variable delay (i.e. $\pi/2$,
$\pi$) for Hahn echo measurements (spin or photon echo), or three
pulses of equal time delay ($\pi/2$, $\pi$, $\pi/2$)  for
projecting Hahn echoes onto population differences, which is
useful when quantum nondemolition readout techniques are available
for measuring populations directly\cite{dandox}.

\begin{figure}
\includegraphics[width=65 mm]{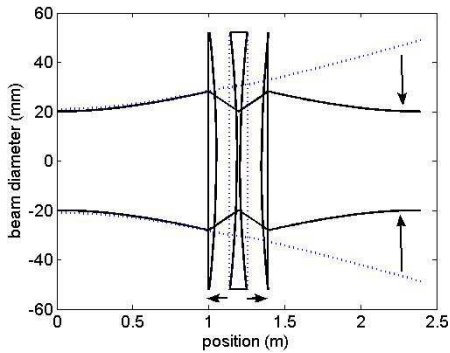}
\caption{Gaussian beam simulation illustrating the change in focal
power with lens separation. Depicted is a system of three lenses
of focal length 0.5 m, -0.25 m, and 0.5 m, respectively (curvature
exaggerated for visibility). A 240 GHz Gaussian beam with an
initial beam waist radius of 2 cm is largely untransformed when
the spacing between the lenses is small (dotted lines). (b) When
the spacing between the lenses is increased to 0.2 m (arrows), the
beam (solid lines) may be symmetrically refocused to its original
beam waist.\label{beamrelay}}
 \end{figure}

The approach taken in the present work is reminiscent of an
optical delay line, with the addition of a set of symmetrically
placed lenses whose relative spacing can be varied to compensate
for diffraction over a wide range of delays. The working principle
is based on the fact that the effective focal power of a composite
lens depends on the spacing between the elements (Fig.
\ref{beamrelay}). Consider a composite lens comprised of a
negative lens of focal length $-f$, flanked by two positive lenses
each with focal length $2f$, which are positioned equal distances
away from the negative lens. Such an arrangement can be used to
relay or ``refocus'' a Gaussian beam. However, when placed
together to form a single block, the composite lens exhibits no
significant transformation on an incident beam. This symmetric
lens arrangement is implemented in our delay line because it
provides diffraction compensation over the widest possible range
of delays, as explained below.

\begin{figure}
\includegraphics[width=70mm]{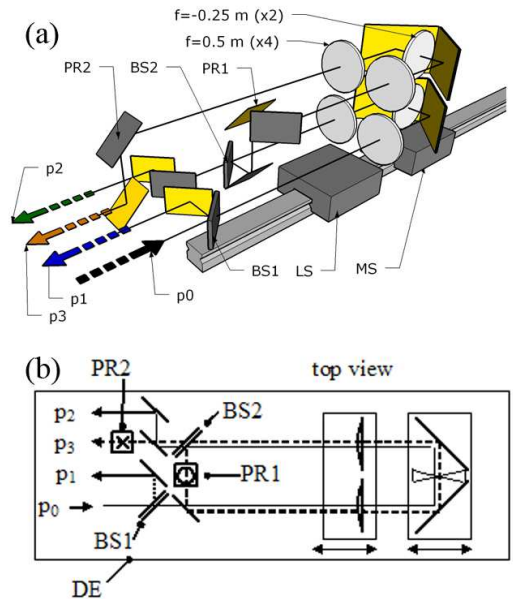}
\caption{Schematic views of a two-level, three pulse,
diffraction-compensating delay line. (a) 3D scaled solid model.
The incident THz pulse (p$_0$) has a beam waist near the first Si
etalon variable beam splitter (BS1). The first (undelayed) exit
pulse (p$_1$) is reflected from BS1. The transmitted portion
follows a delay path of length determined by the location of the
mirror slider (MS) along a sturdy rail. The lens slider (LS) is
adjusted independently to refocus the beam such that a similar
beam waist is formed near the exit. The return beam is split by a
second beamsplitter (BS2); the transmitted portion exits as the
second pulse (p$_2$), while reflected portion (p$_3$) is
periscoped (PR1) up to a second, identical delay path and then
periscoped (PR2) back down near the exit. For scale, the lenses
are 100 mm in diameter. (b) Projected top view schematic diagram
with beam paths offset for clarity. The dotted path represents p1,
the solid path is p$_2$, and dashed path is p$_3$. The entire
delay line is contained within a desiccated enclosure (DE).
\label{schematics}}
 \end{figure}

In the present design (Fig. \ref{schematics}) $45^{\circ}$ bending
mirrors are placed between two plano-convex 0.5 m focal length
lenses and a $-0.25$ m focal length biconvex lens. The mirrors and
negative lens are mounted to a sliding carriage on a precision
rail. The location of the sliding carriage along the rail
determines the length of the delay path. The sliding carriage is
moved via a computer-controlled stepper motor and timing belt with
limit and home switches to forestall collisions and ensure
repeatably referenced positioning. The positive focal length
lenses are mounted pair-wise on a separate computer-controlled
slider. At each delay, the position of the positive lens carriage
can be set to relay to the output a Gaussian beam with a waist
near the input. Other lens arrangements are possible for achieving
the same effect, such as putting two or three lenses in one arm of
the delay. However, the symmetric embodiment was chosen because
(1) it requires only two moving stages, (2) if more rail sliders
were needed for focusing, the additional thickness of the sliders
would limit the minimum path length/delay, and (3) most
importantly for long wavelength radiation, the symmetric design
can be folded back and stacked on top of itself, without the
addition of more moving stages. This is desirable to prevent a
diffracting beam from becoming unmanageably large in the course of
a long delay path

The layout is simultaneously constrained by a number of
experimental requirements. These include a desired small physical
separation of the delayed pulses (which limits the maximum input
beam diameter) and the maximum lens size (which simultaneously
determines the minimum return beam offset and minimum lens
spacing) .These first two requirements (beam input size and
maximum beam diameter) fix the maximum delay that can be
accommodated for a given wavelength. The additional requirements
that a third pulse be twice the delay of the second pulse and that
all outgoing pulses have similar beam waist diameters constrain
the placement of the input beam waist location relative to the two
beamsplitters.

\begin{figure}
\includegraphics[width=60mm]{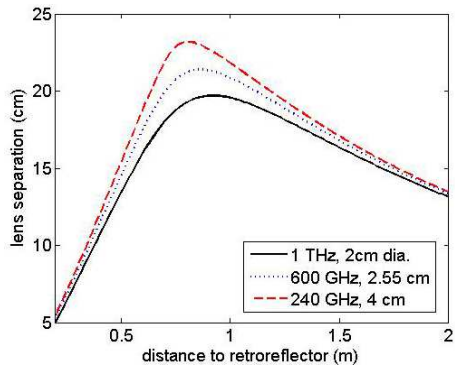}
\caption{Q-parameter calculation of lens separation for relaying a
Gaussian beam. The separation is measured as the distance from the
center negative lens to either positive lens in the configuration
shown in Figure \ref{beamrelay}. Lens focal lengths are 0.5 m and
-0.25 m. Minimum beam waists are chosen for each frequency to
yield real solutions over the entire range of
delay.\label{lenssep}}
 \end{figure}

The optimal lens carriage position for relaying (``focusing'') a
delayed beam is calculated using the ``q'' parameter
method\cite{qparameters}, which is valid in the paraxial
approximation. The thickness of the lenses can be treated within
this formalism but the effect of finite lens thickness was found
to be negligible in the present design. The cross section of the
input beam is assumed to be a Gaussian intensity profile, and is
characterized by two parameters: beam waist radius, and radius of
curvature. The ``focal'' condition requires the outgoing, delayed
beam have the same beam waist radius and radius of curvature as
the incident beam. This is sufficient to determine the required
spacing between the positive and negative lenses. The focal
lengths of the lenses are chosen such that the spacing required to
focus at short delays is not less than the minimum allowed by the
interpositioned bending mirrors. The required lens separation is
then calculated given the delay, the lens focal lengths, and
minimum beam waist radius (Fig. \ref{lenssep}).

The maximum delay per delay line is 12.8 ns (25.6 ns total). The
minimum delay per delay line is limited by the minimum lens
spacing, and is dependent on operating wavelength, falling in the
range 3.5--4 ns. The lenses are kinematically located and can be
removed from the beam path for visible alignment, and for use at
frequencies which do not require diffraction compensation. The
delay line contains two variable beam splitters (Fig.
\ref{schematics}, BS1 and BS2), which can be removed or replaced
with fixed ratio wire mesh beam splitters or fully reflecting
mirrors. A variety of configurations are possible. With the second
beamsplitter removed, the delay line produces two pulses of
relative delay 3.5--12.8 ns. With a mirror at the position of the
beamsplitter, two pulses are delivered with delays of 7-25.6 ns.
Additionally, an extra 3.5 ns compensating path was designed with
diffraction compensating lenses, which can be inserted in the path
of the undelayed beam (Fig \ref{schematics}, $p_1$) to produce
pulse of delay 0--9.3 ns or 3.5-22.1 ns. With the variable second
beamsplitter in place, three pulses are generated, which are
separated by equal, variable delays of 3.5--12.8 ns.

\begin{figure}
\includegraphics[width=60mm]{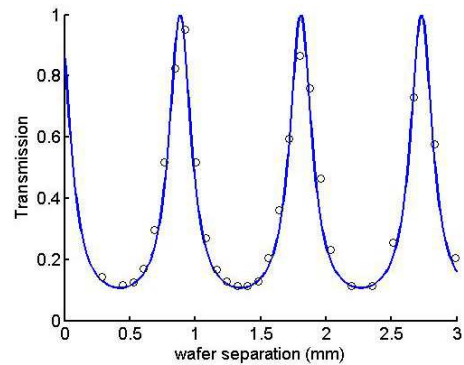}
\caption{Si wafer etalon transmission at 240 GHz. The solid line
represents the theoretical transmission of two 500 $\mu$m thick Si
wafers at 45$^\circ$ angle of incidence, the open circles are
experimental measurements of the beam power normalized to the
power with the wafers removed. The only fit parameter is the
horizontal axis offset. \label{beamsplitter}}
 \end{figure}

For pulses of nanosecond duration, an etalon-based variable beam
splitter is used to control the relative intensities reflected by
the first beamsplitter (Fig. \ref{schematics}, p$_1$) and
transmitted by the second beamsplitter (Fig. \ref{schematics},
p$_3$). The etalon is comprised of two closely spaced 500 $\mu$m
thick Si wafers at 45 degree incidence. The wafer spacing can be
adjusted via a translation stage to achieve transmission in the
range 0.1--1 (reflectivity 0--0.9) (Fig. \ref{beamsplitter}). For
shorter pulses in the tens to hundreds of picoseconds time scale,
fixed beam splitters based on wire meshes should be used, as
multiple reflections from the four interfaces in the etalon would
distort and stretch the pulse.

The lenses are cut from 100 mm diameter polypropylene (PPE) round
stock on a CNC lathe, and the mirrors are fabricated in a
cost-effective manner by e-beam depositing 600 nm of Au on 650
$\mu$m thick polished Si wafers with a 20 nm Ti or Cr adhesion
layer. During operation at frequencies above 0.5 THz, a diaphragm
pump is used to continuously circulate the enclosed 0.5 m$^3$
volume of air through an air dryer/filter, such as is commonly
used in compressed air lines. Atmospheric moisture, as measured by
a NIST traceable hygrometer, is reduced from ambient 40\% to less
than 1\% relative humidity in approximately 45 minutes.

\begin{figure}
\includegraphics[width=70mm]{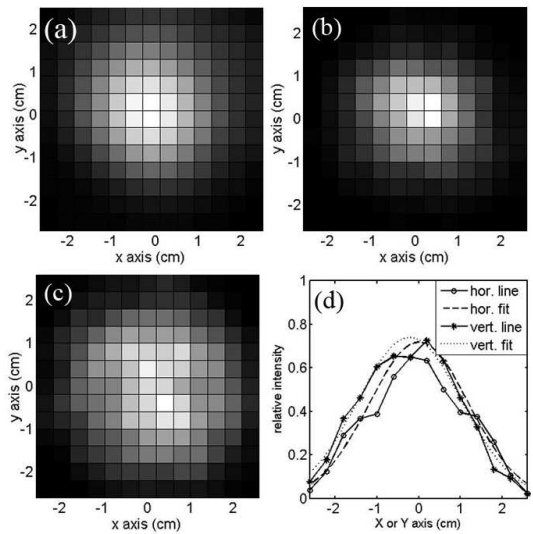}
\caption{Delay line beam profiles at 240 GHz. Beam profiles are
obtained at (a) input, (b) p$_2$ and (c) p$_3$ outputs as a
function of delay. Profiles are measured by rastering a
pyroelectric detector of size 2x2 mm over a 5.2x5.2 cm area. The
slight vertical elongation apparent at long delay settings (e.g.
(c)) is due to a known mirror imperfection. Profiles are shown in
normalized linear grayscale for visibility. (d) For comparison,
centered horizontal and vertical cross sections of figure
\ref{beamprofiles}c are shown with corresponding traces from a
$2D$ Gaussian least squares fit of the form
$a_1\exp{-[((x-x_0)/dx)^2+((y-y_0)/dy)^2]}$, where dx$=2.3$ cm and
dy$=2.4$ cm. \label{beamprofiles}}
 \end{figure}

The delay line is tested with 240 GHz solid state frequency source
based on Schottky diode multipliers that is coupled to a free
space TEM00 Gaussian mode via a corrugated feedhorn. The source
passes through two lenses which match the Gaussian beam to produce
a minimum beam waist diameter of 2 cm at the input of the delay
line (Fig. \ref{schematics}, BS1), which enables us to examine the
effect of unmodeled nonidealities. The most significant
nonidealities include (1) alignment error, (2) input beam
parameter uncertainty, (3) wavefront distortion due to thickness
dependent loss in the PPE lenses, (4) successive clipping of the
beam peripheral fields by mirrors and lenses of size on the order
of the beam waist diameter, (5) the paraxial approximation, and
(6) lens and mirror defects due to material inhomogeneity,
manufacturing defects and strained mounting. The delayed beam is
profiled via a rastering pyroelectric detector of size 2 mm on a
side. The metal housing is carefully shielded with black THz
absorbing foam to prevent feedback on the source. Prior to passing
through two input matching lenses, the beam is indistinguishable
from a perfect Gaussian. Afterward, the center of the beam profile
is slightly flattened due to loss from passing through several
millimeter thick 0.25 m and -0.5 m focal length PPE lenses (Fig.
\ref{beamprofiles}a).

\begin{figure}
\includegraphics[width=75mm]{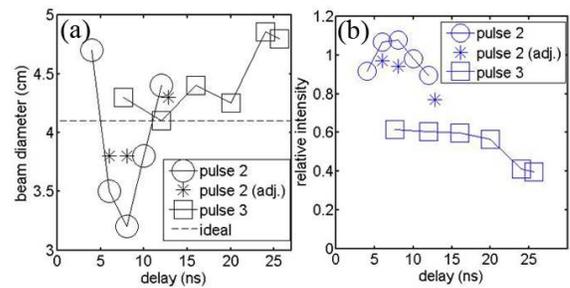}
\caption{Delayed beam properties. (a) Average exit beam waist
diameters ($dx+dy$, from 2D Gaussian fit explained in figure
\ref{beamprofiles}d caption) of a series of beam profiles are
plotted versus delay. Measurement error from repeated measurements
fit within the respective labels. Asterisks label beam waists
measured with the lens separation decreased relative to the ideal
calculated position of figure \ref{lenssep}. (b) Transmitted
power, measured as the integrated beam profile intensity, is
normalized to the power in the $p_2$ output of the delay line at
minimum delay and plotted versus delay. The difference between
p$_3$ and p$_2$ at equivalent delays reflects the insertion loss
($\sim 3$ dB) of the second delay line (including three lenses and
seven mirrors).\label{beampower}}
 \end{figure}

Figure \ref{beamprofiles}b--c show the output beam profiles at
locations p$_2$ and p$_3$ in Figure \ref{schematics}, at short and
long delays, respectively. The beam width of a series of beam
profiles, estimated by a fit to a Gaussian profile (e.g. Fig.
\ref{beamprofiles}d), and plotted in Fig. \ref{beampower}a,
alongside the integrate intensity (Fig. \ref{beampower}b). At
short delays over focusing is noticeable, relative to the expected
result from q-parameter simulations. This is attributed to
distortion of the beam intensity profile by thickness-dependent
loss in the polypropylene lenses. The difference in lens thickness
between the center and edge of the beam is most pronounced in the
positive lenses, where the beam is large. The center part of the
beam experiences more absorption, which results in a flattened
intensity profile that requires less diffraction compensation than
the Gaussian profile assumed by the Q-parameter method. The effect
is only partially compensated by the negative lens, since the beam
diameter is much smaller at the negative lens. At longer delays,
the return beam is slightly wider than expected. This is due to
the far field diffraction from non-Gaussian components (i.e.
M$^2>1$) in the intensity profile introduced by the thickness
dependent lens loss, as well as clipping of peripheral fields by
mirrors of size on the order of the beam diameter, as evidenced by
the negative slope in figure \ref{beampower}a at delay times
corresponding to long delay paths.

The total power in the delayed beams is reduced by a factor of
$\sim$0.5 per delay stage, attributable primarily to reflection
from the six lens surfaces which are not antireflection coated,
secondarily to lens absorption\cite{thzplasticabs}, and lastly to
diffraction losses. While this seems large in comparison to losses
achievable in the optical regime, the utility of a compact,
desiccated delay line becomes apparent when compared to the
typically much higher losses of THz beams traversing the similar
distances in a nondesiccated atmosphere.

\begin{figure}
\includegraphics[width=70mm]{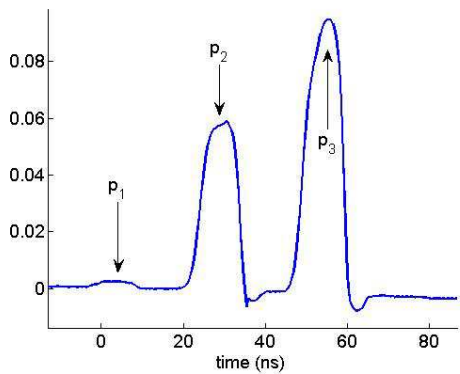}
\caption{Temporal output of delay line at 25.6 ns delay. A 10 ns
duration 240 GHz pulse enters the delay line at p$_0$ in figure
\ref{schematics} and results in three output pulses (Fig.
\ref{schematics}, p$_1$, p$_2$ and p$_3$). The three output pulses
are directed onto a single Schottky diode detector and the
temporal response is recorded by an oscilloscope with 1 GHz
limiting bandwidth. The relative amplitudes of the pulses are
controlled by variable beam splitters (see Fig.
\ref{beamsplitter}), illustrating the wide dynamic range of
reflection and transmission that can be
achieved.\label{threepulse}}
 \end{figure}

The delay of short pulses is demonstrated via use of fast PIN
diode to switch the 15 GHz microwave oscillator that seeds the
multiplier chain of the 240 GHz source, producing single triggered
pulses of 10 ns duration and $\sim$10 mW average power, which are
detected via a Schottky diode (Fig. \ref{threepulse}).

Future quantum control experiments will employ the existing UCSB
free electron laser (FEL) as a frequency-tunable source providing
kilowatts of peak power, together with an optical pulse slicing
system\cite{slicer} which delivers single THz pulses of variable
(3 ps--3 ns) duration at a 1--10 Hz repetition rate. The sliced
FEL pulse will enter the delay line and emerge as two or three THz
pulses of similar size, with variable intensity and delay suitable
for Hahn echo and other pulsed coherent control techniques.

We emphasize that the delay line design presented herein is
optimal for a subset of applications requiring time delayed THz
pulses. There exist a variety of techniques for THz pulse
generation and delay based on the timing of optical pulses,
including photo-activated semiconductor
switches\cite{thzswitch,holearrayswitch}, and THz generation via
ultrafast pulses in nonlinear crystals\cite{leeppln}. When
possible, THz pulse generation based on optical pulse timing is
preferred because optical delay lines do not required diffraction
compensation. In situations where optically gated technologies
cannot produce the correct pulse shape or timing, a number of
methods for physically delaying THz beams can be used. For fixed
delay applications, diffraction in a THz beam can be managed via
waveguides based on polymer, metal or semiconductor wires or
ribbons\cite{siegelwaveguide,grischwaveguide,mittlemanwaveguide}.
For power hungry applications, intolerable waveguide dielectric
losses and insertion loss are the primary concern. In long travel
Michelson interferometers, arrays of corner cube retroreflectors
compensate for diffraction by piecewise retroreflecting the face
of a divergent beam. However, Michelson interferometers back
reflect half the incident power and provide no offset of the
return beam, resulting in a train of decaying pulses, rather than
dividing the power between only two pulses. Deformable mirrors are
another technology which may potentially be leveraged to provide
diffraction compensation.

Future implementations of a diffraction compensating THz delay
line will benefit by using Fresnel lenses, which can be thinner,
and will cause less distortion of the Gaussian beam profile from
thickness-dependent loss. Additionally, for nonideal beams where
diffraction cannot be satisfactorily managed to produce a constant
intensity of pulses versus delay, a method for compensating a
change in power with changes in delay should be implemented. One
such option would be inserting a variable attenuator/reflector in
each of the two delay lines. Another alternative is to use one
variable attenuator and motorize the etalon beamsplitters to
compensate for measured changes in output power. An alternative
layout suitable for 2D NMR-type experiments in which the delay of
a third pulse is varied independently of the second pulse, could
oppose or juxtapose two independent delay lines, with similar beam
splitters and sliding lens carriages. These solutions coupled with
the presented method for diffraction compensation will provide
suitable pulses and delays for coherent control experiments in the
sub-mm and THz spectral ranges.

The authors gratefully acknowledge funding through National
Science Foundation grants NSF-CCF 0507295 and NSF-DMR 0507295.

\bibliography{delaybib}

\end{document}